\documentclass[twocolumn]{aastex631}

\usepackage{tikz}
\usetikzlibrary{shapes.geometric, arrows}
\usepackage{xcolor}
\tikzstyle{startstop} = [rectangle, rounded corners, minimum width=3cm, minimum height=1cm,text centered, draw=black, fill=gray!1]
\tikzstyle{arrow} = [thick,->,>=stealth]
\usepackage{wrapfig}

\graphicspath{{./}{figures/}}
\begin{document}

\title{Estimating Galaxy Parameters with Self-Organizing Maps and the Effect of Missing Data}
\shorttitle{Galaxy properties with SOM}
\shortauthors{La Torre et al.}

\author{Valentina La Torre}
\affiliation{Department of Physics and Astronomy, Tufts University, Medford, MA 02155, USA}
\author{Anna Sajina}
\affiliation{Department of Physics and Astronomy, Tufts University, Medford, MA 02155, USA}
\author{Andy D. Goulding}
\affiliation{Department of Astrophysical Sciences, Princeton University, Princeton, NJ 08544, USA}
\author{Danilo Marchesini}
\affiliation{Department of Physics and Astronomy, Tufts University, Medford, MA 02155, USA}
\author{Rachel Bezanson}
\affiliation{Department of Physics $\&$ Astronomy and PITT PACC, University of Pittsburgh, Pittsburgh, PA 15260, USA}
\author{Alan N. Pearl}
\affiliation{Department of Physics $\&$ Astronomy and PITT PACC, University of Pittsburgh, Pittsburgh, PA 15260, USA}
\author{Laerte Sodré Jr}
\affiliation{Instituto de Astronomia, Geof\'isica e Ci\^encias Atmosf\'ericas, Universidade de S\~ao Paulo, 05508-090, S\~ao Paulo, Brazil}

\shorttitle{Galaxy properties with SOM}

\begin{abstract}

The current and upcoming large data volume galaxy surveys require the use of machine learning techniques to maximize their scientific return. This study explores the use of Self-Organizing Maps (SOMs) to estimate galaxy parameters with a focus on handling cases of missing data and providing realistic probability distribution functions for the parameters.  
We train a SOM with a simulated mass-limited lightcone assuming a $ugrizYJHK_s$+IRAC dataset, mimicking the Hyper Suprime-Cam (HSC) Deep joint dataset.
For parameter estimation, we derive SOM likelihood surfaces considering photometric errors to derive total (statistical and systematic) uncertainties. We explore the effects of missing data including which bands are particular critical to the accuracy of the derived parameters. We demonstrate that the parameter recovery is significantly better when the missing bands are ``filled-in" rather than if they are completely omitted. We propose a practical method for such recovery of missing data.

\end{abstract}

\section{Introduction} 
\label{sec:intro}

The study of the physical mechanisms that drive galaxy formation and evolution \citep[see][for a review]{Somerville_2015} starts with our ability to derive galaxy properties such as their distance, stellar mass, and star formation activity. We also need to do so for large enough galaxy populations that span all potential evolutionary stages, environments and cosmic epochs. 
In the past few decades, increasingly large multi-band imaging and spectroscopic surveys have been constructed to address this need (e.g. SDSS, \citealp{York_2000}; COSMOS, \citealp{Scoville_2007}; CANDELS, \citealp{Grogin_2011}).
Set to start in 2025, the Rubin Observatory Legacy Survey of Space and Time (LSST) will cover 18,000\,deg$^2$ in six filters ($ugrizy$) and reach a 5\,$\sigma$ depth of 27.5 in $r$ \citep{LSST_ref}. In many respects, the three-tiered Subaru Hyper Suprime-Cam (HSC) survey \citep{Aihara_2017} is a precursor of the Rubin LSST survey using nearly all the same filters and reaching comparable depth in its middle (HSC-Deep) tier which covers 27\,deg$^2$.  Large photometric surveys are complemented by spectroscopic surveys (e.g.\,VVDS; \citealp{VVDS_2013}) although spectroscopic samples have been observationally expensive and therefore limited in the past.

Next generation spectrographs will allow for thousands of spectra to be taken simultaneously within wide fields of view allowing for dramatic increases in spectroscopic data samples \citep{Wang2022,PFS,Jin2023}. Survey planning with these upcoming spectrographs, as well as fully contextualizing the results thereof, critically depends on our understanding of the selection functions of the parent photometric samples as well as of the spectroscopic datasets themselves.

Techniques employed for determining galaxy parameters from such multi-band data can be classified as physically motivated or data-driven methods \citep{Salvato19}. The first techniques, already widely used for decades, rely on physical models that have been improved by integrating all the physical knowledge acquired over the years. The estimation is carried out through template Spectral Energy Distribution (SED) fitting to the observed photometry. This method helps us derive parameters such $z_{\rm phot}$, stellar mass, stellar population age, metallicity, star formation history, etc. \citep{Gallazzi_2005}. 
This approach works because each physical parameters leaves specific marks on the SED, determining its shape and amplitude.
Template libraries, spanning a range of physical parameters, are constructed ahead of time or on the go in the process of exploring the parameter space. 
Then, stellar population parameters are assigned to the galaxy based on the properties corresponding to the SED model that best fits the observed photometry 
\citep[e.g.,][]{Sawicki_1998,Walcher_2010,Conroy_2013,Leja_2017}. 
The more sophisticated, and hence realistic, the modeling (e.g.\,treating the Star Formation History SFH as non-parametric in \citealp{Leja_2017}), the more computationally expensive the SED fitting becomes. Multiplying this by millions and soon billions of galaxies makes this traditional approach largely unfeasible. The necessity of speeding up the process of galaxy parameter estimation from observed photometry or spectroscopy is one of the reasons for the widespread applications of machine learning ML in astronomy \citep{Ball2010,Acquaviva_2015,Baron19}. Machine learning methods are data-driven since they learn directly from the data used to train the algorithm. Techniques include data compression, dimensionality reduction, visualization techniques, etc. which all help to manage and process large amounts of information. Machine learning methods can both offer significant speed-up over traditional SED fitting, but also as they learn complex connections amongst multi-dimensional data can help us make additional inferences from them. 

As stated above, ML is already widely used in astrophysics and cosmology \citep{Longo_2019, Baron19}. \cite{Salvato19} provides a comprehensive review of different ML methods that can be applied to estimate $z_{\rm phot}$. Star Formation Rate SFR and stellar mass $M_*$ are predicted by training deep neural networks for the Galaxy And Mass Assembly (GAMA) survey \citep{Surana_2020}. \cite{Lovell_2019} derive SFHs by training convolutional neural networks (CNN) with simulated galaxy spectra from Illustris and EAGLE. \cite{Acquaviva_2015} applies a range of supervised ML methods (regularized ridge regression, RF, extremely randomized trees ERT, boosted decision trees AdaBoost and support vector machines) to predict galaxy metallicty from five-band SDSS photometry. The Cosmology and Astrophysics with MachinE Learning Simulations (\textsc{CAMELS}) project \citep{Villaescusa_Navarro_2021} was created to offer a large, state-of-the-art hydrodynamic simulation that can serve as a training dataset for a variety of ML methods. One such application involved using neural networks to estimate cosmological and astrophysical parameters from 17 simulated galaxy properties, revealing significant correlations between $\Omega_m$ and galaxy properties \citep{Villaescusa_Navarro_2022}.

The Self-Organizing Map \citep[SOM,][]{kohonen1982} is an unsupervised ML algorithm that 
projects high dimensional data onto a 2D map that nevertheless preserves the topology. In the case of galaxy surveys, inputting for example a catalog of 10$+$ observed colors results in galaxies with similar colors being grouped closely on the 2D map. This makes SOMs a powerful tool for visualizing large astronomical surveys \citep{Geach_2012,Longo_2019}. Because observed colors depends on the redshift and stellar population parameters of galaxies, this effective grouping by multi-dimensional colors can also be used for accurate and computationally very efficient derivation of said parameters. SOMs have been used in the literature to classify stellar spectra \citep{SOM_Stellar_Spectra}, classify galaxy morphology \citep{Galvin_2019}, estimate photometric redshifts and other galaxy parameters \citep{Masters_2015,Hemmati_2019,D19,Davidzon_2022}, or for target selection (to have spectroscopic samples that are more representative of the photometric ones) \citep{Hemmati_2019,Masters_2019}.

The goal of this paper is to investigate the use of SOM for parameter estimation including the effects of photometric errors and missing data, a common scenario in real astronomical datasets that is not adequately handled within SOM applications. To achieve this, we employ a fiducial $K_s$-magnitude limited sample to mimic the joint HSC-Deep photometric dataset. This analysis lays the groundwork for future work involving training a SOM with real data from the HSC-Deep+ survey.

This paper is organized as follows. The employed simulated mass-limited and magnitude-limited datasets are described in detail in Section\,\ref{sec:dataset}. In Section\,\ref{sec:method} we describe the set-up, training, labeling and visualization of our SOM. In Section\,\ref{subsec:visualization_obs} we present the visualization of the observed-like sample and the lessons learned therefrom. In Sections\,\ref{subsection:parameter_estimation}-\ref{sec:upper limits} we present parameter estimation results with further investigation on how derivations are influenced by realistic cases like noise and gaps in the photometric coverage. In Section\,\ref{sec:discussion} we discuss the implications of our findings as well as compare them with the literature. In Section\,\ref{sec:conclusions} we present our summary and conclusions. In Appendix A, we compare a SOM trained on a simulated vs. an observed dataset. 
Throughout this paper, all the magnitudes are expressed in the AB system \citep{Oke_Gunn_1983}, and we adopt $\Lambda$CDM cosmology with H$_0$ = 70 km s$^{-1}$ Mpc$^{-1}$, $\Omega_{M}$ = 0.3, and $\Omega_{\Lambda}$ = 0.7.  

\section{Data} \label{sec:dataset}
We start with a mass-limited simulated dataset which we color-calibrate with real galaxy data to provide realistic color-parameter correlations. We then generate a fiducial magnitude-limited simulated dataset which is modelled after the HSC-Deep survey with ancillary data.
Below we present the details of our mass-limited initial dataset, its color-calibrations and finally the observations-like simulated dataset.

\subsection{Simulated lightcone \label{subsec:masslimited_sample}}

Our analysis is based on two publicly-available mock catalogs \citep[described in ][]{Laigle2019} with observed-frame photometry extracted from the Horizon-AGN light-cone\footnote{\url{https://www.horizon-simulation.org}} \citep{Dubois14}. Horizon-AGN is a cosmological hydrodynamic simulation that contains 1024$^3$ dark matter particles in a box of $L_{box} = 100$ h$^{-1}$Mpc, corresponding to a dark matter mass resolution of $8 \times 10^7$M$_{\odot}$. The simulation is run with the adaptive mesh refinement code \textsc{ramses} \citep{ramses}, following the evolution of the gas including the effects of gravity, hydrodynamics, gas cooling and heating processes, star formation and stellar feedback, and feedback from black holes \citep[see][for details]{Kaviraj_2017}. Gas is heated from an uniform UV background after the reionization epoch, according to \cite{Haardt_1996}, and it cools down to $10^4$\,K via H, He and metals following \cite{sutherland_1993}. Star formation is activated only in regions with gas number density $n > 0.1 $ H cm$^{-3}$ and according to the Schmidt law \citep{Schmidt_1959}: $\dot \rho_{SFR} = \epsilon \rho_{gas} / \tau_{ff}$, where  $\dot \rho_{SFR}$ is the SFR mass density, $\epsilon$ is the star formation efficiency, $\rho_{gas} $ is the gas mass density, and $\tau_{ff}$ is the free fall time of the collapsing gas cloud. Stellar feedback is provided by stellar winds and type Ia and II supernovae, and AGN feedback is a combination of the radio or quasar mode based on the black hole accretion rate.

In \cite{Laigle2019}, the 1\,deg$^2$ light-cone is build by running the \textsc{AdaptaHOP} halo finder \citep{10.1111/j.1365-2966.2004.07883.x} on the Horizon-AGN light-cone over $0 < z < 4$ and selecting structures with a density threshold 178 times the average matter density at that redshift, and then considering the total stellar mass within each halo. The extracted light-cone contains 789,354 galaxies with stellar masses $>$10$^9$M$_{\odot}$ in a redshift range $z=0 - 4$. For each galaxy, \citet{Laigle2019} compute its SED by adopting single stellar population models from \cite{Bruzual_Charlot} and a Chabrier IMF \citep{Chabrier}. They adopt an empirical conversion relation from the gas phase metals to dust and estimate the spatial distribution of that dust from the gas density in each cell. Integrated along the line of sight, this provides the column density of dust. Dust attenuation is then applied to the spectra by assuming the dust follows the R$_V = 3.1$ Milky Way dust model by \cite{Weingartner_2001}. They do not take into account scattering in or out of the line of sight, but only dust absorption. This results for example in higher (by $\approx$0.8dex) estimated magnitudes in the UV \citep{Laigle2019}. They also account for absorption by the IGM, but not for foreground extinction by the Milky Way, which is expected to be corrected for in any observed survey that would be compared with this simulated dataset.  

\begin{figure}%[H]
\centering
\includegraphics[width=0.2\textwidth, trim={8cm 6.5cm 8cm 6.5cm}]{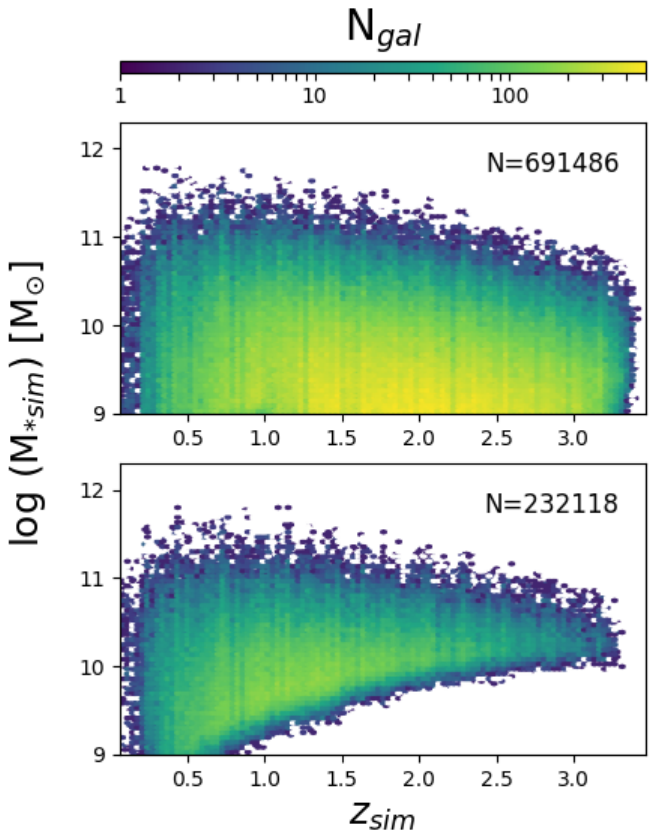}
\caption{{\it Top: }The density of galaxies in the stellar mass versus redshift plane from the mass-limited 1\,deg$^2$ simulated Horizon-AGN light-cone, after applying a $u$-magnitude cut of $m_u < 28$ (Sec.\,\ref{subsec:masslimited_sample}). {\it Bottom: } The same simulated dataset after color-calibration and applying a $K_s$-magnitude cut of $K_s<23.5$ (Sec.\,\ref{subsec:obs_samples}).}
\label{figure1}
\end{figure}

Apparent magnitudes are obtained by convolving the thus computed galaxy spectra with the desired filter profiles. 
Of the two mock catalogs\footnote{\url{https://www.horizon-simulation.org/data.html}} we use, one is the COSMOS-like mock where the photometry matches the \cite{Laigle2016} COSMOS2015 catalog. Specifically, this catalog includes the \textit{u}$^*$-band from MegaCam on the Canada-Hawaii-France Telescope CHFT \citep{MegCam}, 6 broad optical bands (\textit{B}, \textit{V}, \textit{r}, \textit{i}$^+$, \textit{z}$^{++}$), 12 median and 2 narrow bands from the Suprime-Cam on the Subaru telescope \citep{Taniguchi_2007, Taniguchi_2015}, and the \textit{Y}-band from the Hyper Suprime-Cam/Subaru \citep{HSC_instrument}, which is slightly bluer than the $Y$ filter from VIRCAM. The COSMOS2015 also includes near-IR \textit{YJHK}$_s$-band from VIRCAM on the VISTA telescope \citep{VIRCAM, UvistaDR2}, and the \textit{H} and \textit{K}$_s$ bands from WIRCam/CFHT \citep{WIRCam, McCracken_2010}. Finally, the catalog includes the mid-IR {\sl Spitzer} IRAC channels centred at 3.6 and 4.5$\mu m$ (\textit{ch1} and \textit{ch2} respectively) \citep{Fazio2004}.  
For our analysis, we need a simulated mock of the HSC-Deep survey \citep{Aihara_2017} and its ancillary surveys as described in more detail in Section\,\ref{subsec:obs_samples}. To provide a similar photometric coverage, we use the $u$, $r$, $i$, $z$, $Y$, $J$, $H$, $K_s$, $ch1$ and $ch2$ photometric datapoints from the Horizon-AGN COSMOS-like mock. However, the COSMOS-like mock is missing the $g$-band. So we also use the public \textit{Euclid}+LSST-like mock which includes the LSST $g$-filter which is very close to the HSC $g$-filter. From these simulated photometric catalogs, we use only the photometry without any photometric errors applied since we add photometric errors ourselves for the specific surveys we are considering (see Section\,\ref{subsec:phot_err}). In cross-matching the two mocks $\approx100$ objects are lost leaving us with 789,292 galaxies. In the simulation, the \textit{u}-band exhibits a tail that extends to highly unrealistic values (corresponding to this band dropping shortward of the Lyman break). To mitigate this issue, we consider only galaxies with $m_u < 28$, leaving us with 691,486 objects. This cut limits the redshifts to $z\sim3.4$, as shown in Figure\,\ref{figure1} (\textit{top},) but has no other effects on the dataset.

\subsubsection{Color-calibration}
\label{subsec:color-calibration}
We expect the magnitude and color distributions of this light-cone to not be quite right, because of the simplicity of the dust attenuation treatment, as well as the fact that this idealized photometry does not take into account all the systematic effects that one would have from extracting photometry from an image \citep{Laigle2019}. We correct for this systematic offsets in the photometry following the method of \citet{pearl22}. This method adjusts simulated magnitudes to match the correlation between $M/L_{\nu, obs}$ ratios and sSFR$_{UV, obs}$, ensuring consistency between simulations and observations.

We performed the color-calibration using the UltraVISTA catalog (UVISTA; \citealp{muzzin13}) which has similar photometric coverage as well as all necessary derived quantities including redshift. To ensure maximum reliability for our color-calibration, we selected galaxies from the UltraVISTA catalog based on the following criteria: 0.2 $<$ \texttt{z$\_$peak} $<$ 4.0, 5 $< logM_* <$12 [M$_{\odot}$], \texttt{star}=0, \texttt{contamination}=0, \texttt{nan$\_$contam} $<$ 3, \texttt{use}=1, \textit{K$_s$} $<$ 23.5 AB. This results in a K$_s$-limited catalog with only $\sim$100 objects at $z>3$.

We derived the SFR$_{UV}$ for both the simulated dataset and UVISTA following the \cite{k12} conversion relation from $L_{NUV}$, which was available for both datasets. For the sake of the calibration, we excluded simulated galaxies with log(sSFR$_{UV}$)$<$-13 and log(sSFR)$<$-13, as such values are not typically observed. 

\begin{figure}%[H]
\centering
        \includegraphics[width=0.45\textwidth,trim={0 0 0 0}]{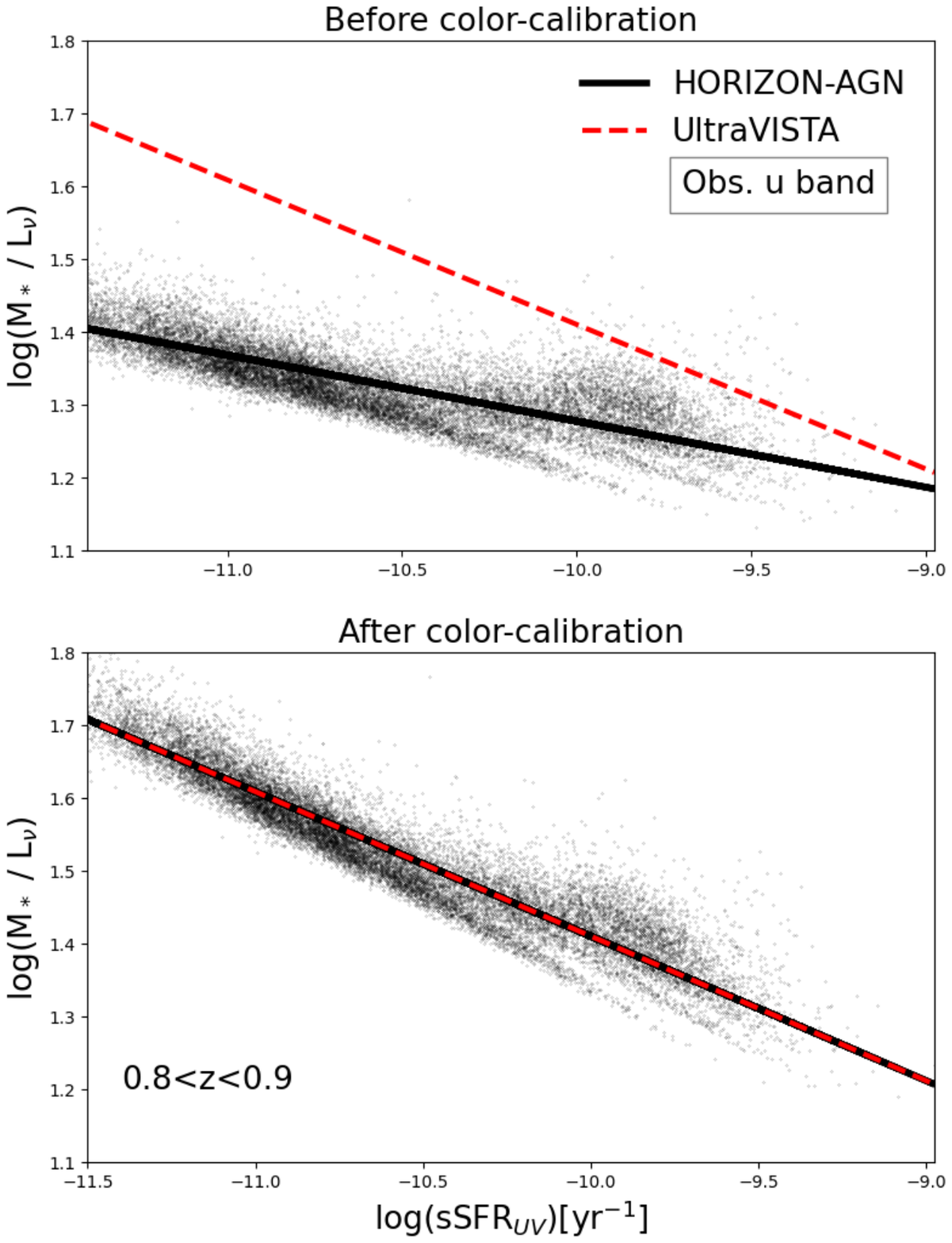}
\caption{$M/L_{\nu,u}$ ratio for the observed \textit{u}-band vs. the sSFR$_{UV}$ in the redshift bin $0.8 < z < 0.9$, before and after color-calibration in the \textit{upper} and \textit{lower} panel, respectively. The black data points and their relative best-fit (black solid line) represent the simulations. The red dashed line represents the best fit to the observed UVISTA catalog \citep{muzzin13}. These best-fits are used to adjust the simulated photometry to match the observed relation. We perform such color-calibration for all bands in all redshift bins.} 
\label{figure2}
\end{figure}

We perform the color calibration in 70 redshift bins. The first 69 bins cover the range $0<z\leq2.5$, each with $\approx$8K simulated galaxies, making $k$-correction negligible within these narrow redshift bins. The remaining 132,281 simulated galaxies at $z>2.5$ are all grouped in the final bin. The choice of this binning is motivated by the use of the $K_s$-selected UVISTA catalog for the calibration, implicating that there are few observed objects at $z>2.5$. Consequently, dividing this redshift range into smaller bins would yield too few objects for a reliable fit in the log(M/L$_{\nu}$)-log(sSFR$_{UV}$) plane.
Figure\,\ref{figure2} shows as an example the $M/L_{\nu,u}$ ratio for the \textit{u}-band vs.\,the sSFR$_{UV}$ for the redshift bin $0.8 < z < 0.9$. We chose as an example the $u$-band as the needed calibration is by far the strongest in this band. 
As shown in Figure\,\ref{figure2}, prior to color-calibration, the slopes of the fitted lines in the log(M$_*$/L$_{u}$)-log(sSFR$_{UV}$) plane differ between the Horizon-AGN mock and UVISTA data. To align the simulated fitted line with the observations, we perform linear fits to both distributions and calculate $\Delta_{m}$ (slope difference) and $\Delta_b$ (y-intercept difference). The new luminosity in each band and redshift bin is derived as:

\begin{equation}
    L_{\nu,new} = \frac{L_{\nu,old}}{10^{{\Delta_b}} \times sSFR_{UV}^{\Delta_{m}}}
\end{equation}
where $L_{\nu,old}$ is the old luminosity in a specific band $\nu$.
We then derive new apparent magnitudes from:
\begin{equation}
m_{\nu,new} = m_{\nu,old} - 2.5\times log(L_{\nu,new}/L_{\nu,old})
\end{equation}

In this way we obtain the \textit{color-calibrated} version of the mock dataset. For the rest of this paper we refer to this mass-limited, color-calibrated mock as the \textit{mass-limited simulated dataset}, and it is what we use to train the SOM (see Section\,\ref{sec:method}) as well as to generate the observations-like sample as described in the next sub-section.

\subsection{The $K_s$-limited dataset}
\label{subsec:obs_samples}
From the mass-limited simulated dataset described above, we construct a simple $K_s$-limited sample that is meant to 
mimic the Hyper Supreme-Cam (HSC) Deep survey and ancillary data.

The HSC Deep survey is one of the three layers of the HSC survey \citep{Aihara_2017}. The HSC is a 1.8\,deg$^2$ field of view imaging camera \citep{Miyazaki_2018} mounted on the 8.2-m Subaru telescope on the summit of Mauna Kea in Hawaii, operated by the National Astronomical Observatory of Japan (NAOJ). The product of its wide field of view and the large collecting area of the telescope makes HSC second only to the anticipated LSST Camera \citep{Aihara_2017}. 
The Deep layer, which is our focus, covers 27\,deg$^2$ spread across four well-studied extragalactic fields (XMM-LSS, COSMOS, ELAIS$\_$N1, and DEEP2-F3). The HSC photometry includes five broad bands (\textit{g, r, i, z, Y}) and three narrow-band filters (NB387, NB816, NB921). 

The HSC Deep joint dataset augments the above through the addition of several complementary surveys. The $u$-band coverage is provided by the CLAUDS survey \citep{Sawicki19}, which uses the Megaprime camera on the CFHT telescope. The near-IR $J$, $H$, and $K_s$ bands are supplied from a variety of surveys including  UKIDSS/DXS $\&$ UDS \citep{UKIDSS}, VIDEO \citep{Jarvis13}, UltraVISTA \citep{muzzin13}, and DUNES\footnote{\url{http://gxn.as.arizona.edu/DUNES/}} (in prep). \textit{Spitzer} IRAC imaging in channels 1 and 2 (3.6 and 4.5 $\mu$m) is supplied from the DeepDrill survey \citep{Lacy_2020}, the SHIRAZ survey \citep{Annunziatella2023} as well as a number of surveys within COSMOS \citep[e.g.][]{Sanders_2007}. 

At the time of writing this paper, the precise properties of the multi-band photometric catalog for this HSC-Deep joint dataset had not been determined. To test the degree to which our HSC-like mock catalog depends on the precise catalog construction method, we explored two extremes in such a catalog construction. The first extreme, which results in the smallest number of galaxies included, required a 5\,$\sigma$ detection in each band from the $u$-band through to IRAC $ch2$. Our second extreme simulates a scenario where one uses a reference image for source detection and performs forced photometry on all other images, thus producing a photometry value regardless of whether the brightness of any given source exceed 5\,$\sigma$ in these other bands or not \citep[e.g.][]{UvistaDR2}. This method would result in simpler selection function (a single magnitude cut) and a larger number of galaxies than the first method. For this method, we chose to make a simple $K_s$ cut of $K_s<23.5$ without any additional cuts. We find that $75\%$ of the $K_s$-limited sample are also within the first HSC-like mock where we required a 5\,$\sigma$ cut in all bands. Thus, for the bulk of the galaxies our results are independent of the specifics of the catalog construction. While we ran our analysis on both HSC-like mocks, we only show in this paper the results of the runs with the $K_s$-limited mock because of its more extended redshift and mass coverage. There was no significant difference in our findings on parameter estimation accuracy or effects of missing data between the two HSC-like mocks. Our adopted HSC-like $K_s$-limited dataset contains 232,118 galaxies, and its stellar mass vs. redshift distribution is shown in the bottom panel of Figure\,\ref{figure1}.

The bottom panel of Figure\,\ref{figure1} shows that the $K_s$-magnitude cut introduces an incompleteness in stellar mass as a function of redshift, with the stellar mass limit increasing with increasing redshift. As a result, low-mass galaxies are excluded from the sample at $z\gtrsim1$, in particular those with log($M_*/M_{\odot}$)$<10$ are entirely absent by $z\sim3$. We caution that the simulated lightcone we started with, represents 1\,deg$^2$ whereas the HSC-Deep survey covers 27\,deg$^2$. Rarer populations such as log($M_*/M_{\odot}$)$>11$ galaxies at $z>3$ can be present in the real data even though largely missing here.

\section{METHOD}
\label{sec:method}

\subsection{Self-organizing maps}

A Self-Organizing Map \citep[SOM;][]{kohonen1982} is an unsupervised artificial neural network that performs a dimensional reduction of a multi-dimensional parameter space to a lower-dimensional space, while preserving the topology of the data. It thus consists of a lower-dimensional grid where each pixel is characterized by a weight vector that represents the mapping from the higher-dimensional to the lower-dimensional space. The training involves determining these ``weights" so that objects that are similar in the high-dimensional parameter space are grouped together, maintaining the intrinsic structure of input data. For this reason, SOM is a powerful visualization tool especially for a final 2D grid, as considered in this work.
We use the \texttt{Python} library \textsc{SomPY}\footnote{\url{https://github.com/sevamoo/SOMPY}} \citep{moosavi2014sompy} for constructing and training our SOM. Below we describe in more detail how the algorithm works.

The first step is selecting a representative training data, which is then normalized by \textsc{SomPY} to the unit variance with mean of zero. The weight vector of each SOM pixel, or neural network, in \textsc{SomPY} is initialized using principal component analysis \citep[PCA;][]{PCA} to bring the initial weights closer to the input data (\textsc{SomPY} also allows for random initialization, but it extends the training duration). At this point, the Euclidean distance between each neurons and input data is calculated. The neuron with the smallest distance to an input data becomes the \textit{Best Match Unit} (BMU). The latter is thus dragged closer to the data point and neurons around the BMU, namely within a `neighborhood radius', are also dragged in the same direction. This process is repeated in an iterative way until all neurons are as close as possible to the input data. This process is unsupervised as it does not require any a \textit{priori} labelling of input data, unlike for example traditional neural networks where the weights are optimized to match some particular outputs vector. After this optimization, similar objects in the high-dimensional parameter space will be grouped together in the final lower-dimensional (usually 2D) space, maintaining the intrinsic topology of input data.

\begin{figure*}
\centering
    \includegraphics[trim={3.4cm 9cm 2.1cm 5cm},width=1\textwidth]{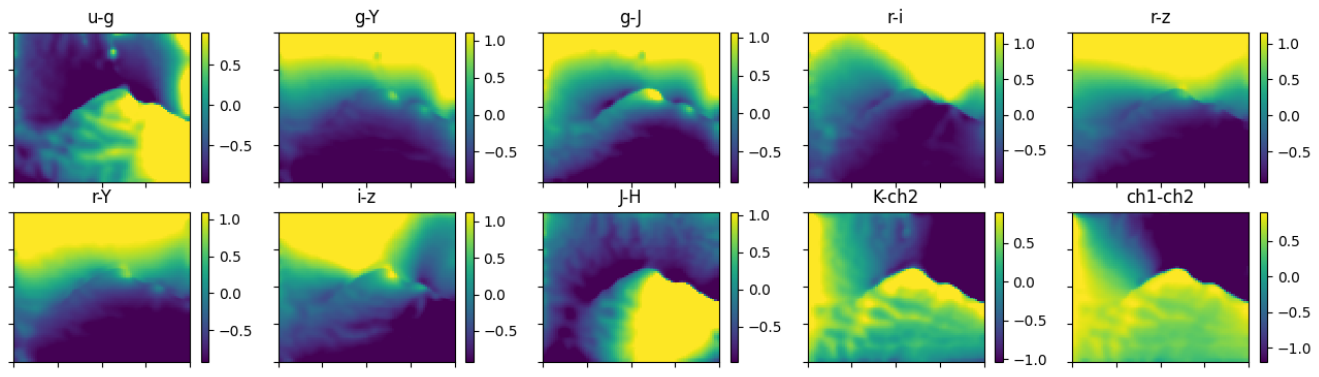}
    \caption{Component maps showing each of the 10 input normalized colors across our trained SOM. The color bar indicates the normalized color values.}
\label{figure3}
\end{figure*}

\begin{figure*}
\centering
    \includegraphics[width=0.75\textwidth,trim={0cm 0.9cm 0cm 0cm}]{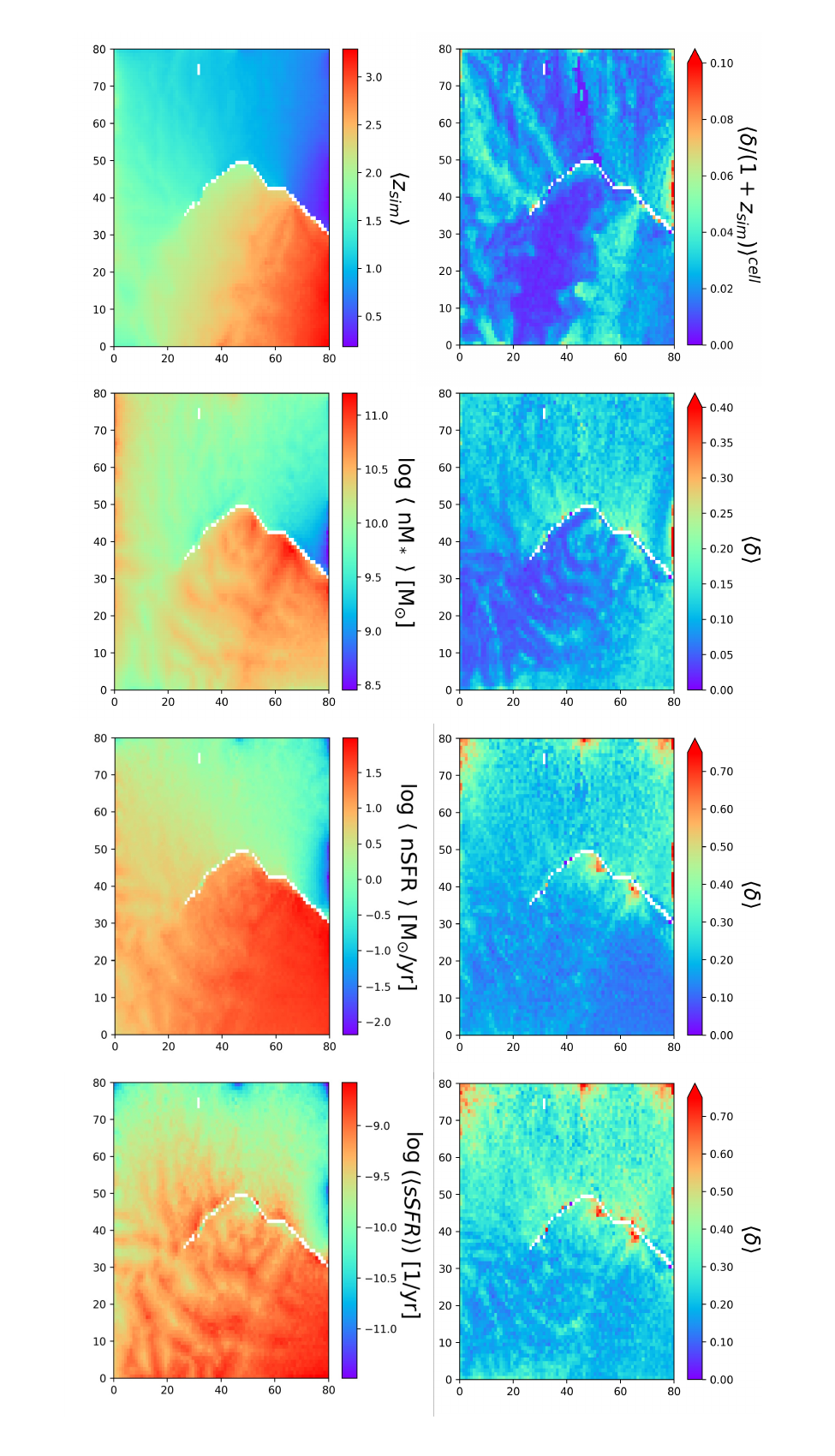}
    \caption{\textit{Left panels}: SOM with cells color-coded according to the median of galaxy redshift, normalised stellar mass, normalised SFR, and specific SFR respectively from the top to the bottom panel. \textit{Right panels}: SOM with cells color-coded according to the distribution width $\langle\delta\rangle=\langle84\%-16\%\rangle$ of the corresponding galaxy parameters on the left side.}
    \label{figure4}
\end{figure*}

\subsection{The SOM trained with galaxy colors}
\label{subsec:idealSOM}

In this work, the input parameter space has ten dimensions representing observed-frame colors for galaxies. These colors, based on the expected photometry available for the HSC-Deep joint dataset (see Section\,\ref{subsec:obs_samples}) are: $u-g$, $g-Y$, $g-J$, $r-i$, $r-z$, $r-Y$, $i-z$, $J-H$, $K-ch2$, $ch1-ch2$. We select these colors for their sensitivity to galaxy parameters such as redshift, normalized $M_*$ and SFR, and sSFR. To do so, we computed the distance correlation for all possible two-color combinations (55 in total) between the 11 available bands. This analysis identified the 10 colors with the strongest non-linear relationships with our target parameters. While we initially explored using adjacent colors for SOM training, our final selection minimized parameter spread within SOM pixels (Sec.\,\ref{subsec:acrossSOM}), leading to more accurate parameter estimations, particularly for $M_*$ and SFR.

For our training set, we selected a random subset from the mass-limited dataset. The training set consists of 228,524 galaxies or about 1/3 of the total. The performance of SOM depends on the user set map's size and geometry which need to be verified by the user depending on their specific case \citep[see][]{D19}. The choice of size and geometry of the map, should represent a compromise between good sampling of the data and a high resolution (less quantization error). We achieve this, following the \citet{D19} procedure, by training different SOMs starting with a square 20x20 map and gradually increasing the size by adding 10 cells in both dimensions. We kept track of sampling vs.\,error finding that an 80$\times$80 SOM represents a good compromise, similar with the \citet{D19} findings. We also tested the effect of different aspect ratios, but found that a square map gave the best performance. Finally, in our final 80$\times$80 SOM, we verified that the 10 color distributions in the input dataset and across SOM pixels are consistent. This ensures that thus trained SOM is representative of the input data\footnote{For this test, one needs to denormalize the SOM component pixel values since \textsc{SomPY} normalizes the input data by the variance in order to keep the data ranges comparable.}. 

\subsection{Visualization}
\label{subsec:visualization}

Figure\,\ref{figure3} shows the distribution of each of the 10 normalized colors across the map. The distribution is typically smooth except for a sharp discontinuity in the middle of the SOM, more evident for some of the colors such as $u-g$ or $ch1-ch2$. This is because, in this case, the algorithm organized the input data in a ring-like structure, which causes the colors to change in a counter-clockwise direction. The point where the loop closes marks the discontinuity. However, this discontinuity is not random, but rather it has a physical meaning, denoting the transition between galaxies with low and high redshift and other stellar population properties. For instance, for $ch1$ - $ch2$ we expect redder colors at $z > 1.7$ when both IRAC channels sample blueward of the 1.6\,$\mu$m bump. This means that lower values of $ch1 - ch2$ on the map would correspond to $z < 1.7$ and vice versa. The $u-g$ colors also switch from low to high across the same discontinuity, likely due to the effects of redshift and SFR. Lower redshifts correspond to lower-mass, passive galaxies, while higher redshifts correspond to redder colors, indicating more massive, star-forming galaxies.

At $z>1.7$ the $u-g$ colors redden further as a result of the Lyman break, which enters the $u$ filter at $z>2.2$. In general, both redshift and galaxy stellar population properties affect these distributions. Figure\,\ref{figure3} clearly demonstrates that galaxies with similar 10-color sets are grouped together in the trained SOM. 

Considering our ultimate goal is to replicate the methodology presented in this work with observed data, we conduct an additional sanity check shown in Appendix\,\ref{appA}. This check involves verifying the similarity of the component maps of two different SOMs, one trained with an observed dataset (UVISTA) and the other with simulated data where we match as closely as possible the selection function of the observed dataset. We find encouragingly consistent component maps. %Further details regarding this test can be found in Appendix\,\ref{appA}.

\subsection{SOM pixel labeling} 
\label{subsec:acrossSOM}
The color distributions across the SOM (Figure\,\ref{figure3}) already allowed us to identify trends with redshift and stellar population parameters. 
To quantify these trends and use SOM for parameter estimation, we need to label each SOM pixel with the galaxy properties based on its corresponding 10-dimensional color combination. There are multiple approaches. \cite{Masters_2015} use the sub-set of their galaxies with spectroscopic redshifts to label each pixel by the expected redshift. \cite{Hemmati_2019} use simple empirical template libraries covering a range of parameters such as age of the stellar population and dustiness. In our case, following the approach of \citet{D19}, we label SOM pixels with the median of each parameter. The latter are the Horizon-AGN parameters from the training set galaxies falling within a given pixel. This works for redshifts and sSFRs which only depend on the observed-frame colors. Stellar mass M$_*$ and SFR, however, require information about the overall amplitude of a galaxy's spectrum, which is not modeled in our colors-trained SOM. To account for this amplitude-dependence, we normalise M$_*$ and SFR values from our simulated dataset to a reference magnitude $J$ = 23 mag by doing: 
\begin{equation}
	nX = X \times 10^{-0.4 \times (23.0 - J)},    
    \label{eq:normalization}
\end{equation}

where $X$ is the galaxy parameter that needs to be normalized and $J$ is the $J$-magnitude for each particular galaxy. 

The left panels of Figure\,\ref{figure4} show how redshift ($z$), normalised stellar mass (nM$_*$), normalised star formation rate (nSFR), and sSFR vary across the SOM. Consistent with our initial visual inspection, the considered parameters increase in a counter clockwise trend. 
%redshift shows a rough left-to-right trend whereas sSFR shows an up-down trend. 
Specifically, more distant galaxies at $z > 2.5$ are also more massive and active. Note that the SOM's morphology is not universal; it varies depending on the specific colors and dataset used for training. However, its ability to reveal trends remains consistent. For instance, comparing these plots with Figure \ref{figure3}, it is evident that redder galaxies are the farther, more massive ones. 

The right panels of Figure\,\ref{figure4} show the distribution width $\delta$ for each parameter. The width is quantified by the difference between the 84th and 16th percentiles among galaxies within each pixel. The $\delta/2$ represents the systematic uncertainties $\sigma_{sys}$ of the derived redshifts and stellar population pararameters using SOMs. Large parameter spread within a pixel can be due to degeneracies, although the use of 10 different colors tends to mitigate this to a large degree. We also notice that large spreads in our SOM tend to concentrate the edges. This could be due to: 1) boundary effects, as \cite{D19} explains, risen from galaxies with extreme colors put on the borders of the map by the algorithm; 2) poorly defined areas, where there are not enough objects to train properly a cell.

\begin{figure}%[H]
\centering    
\includegraphics[width=0.5\textwidth,trim={3cm 0 0 0}]{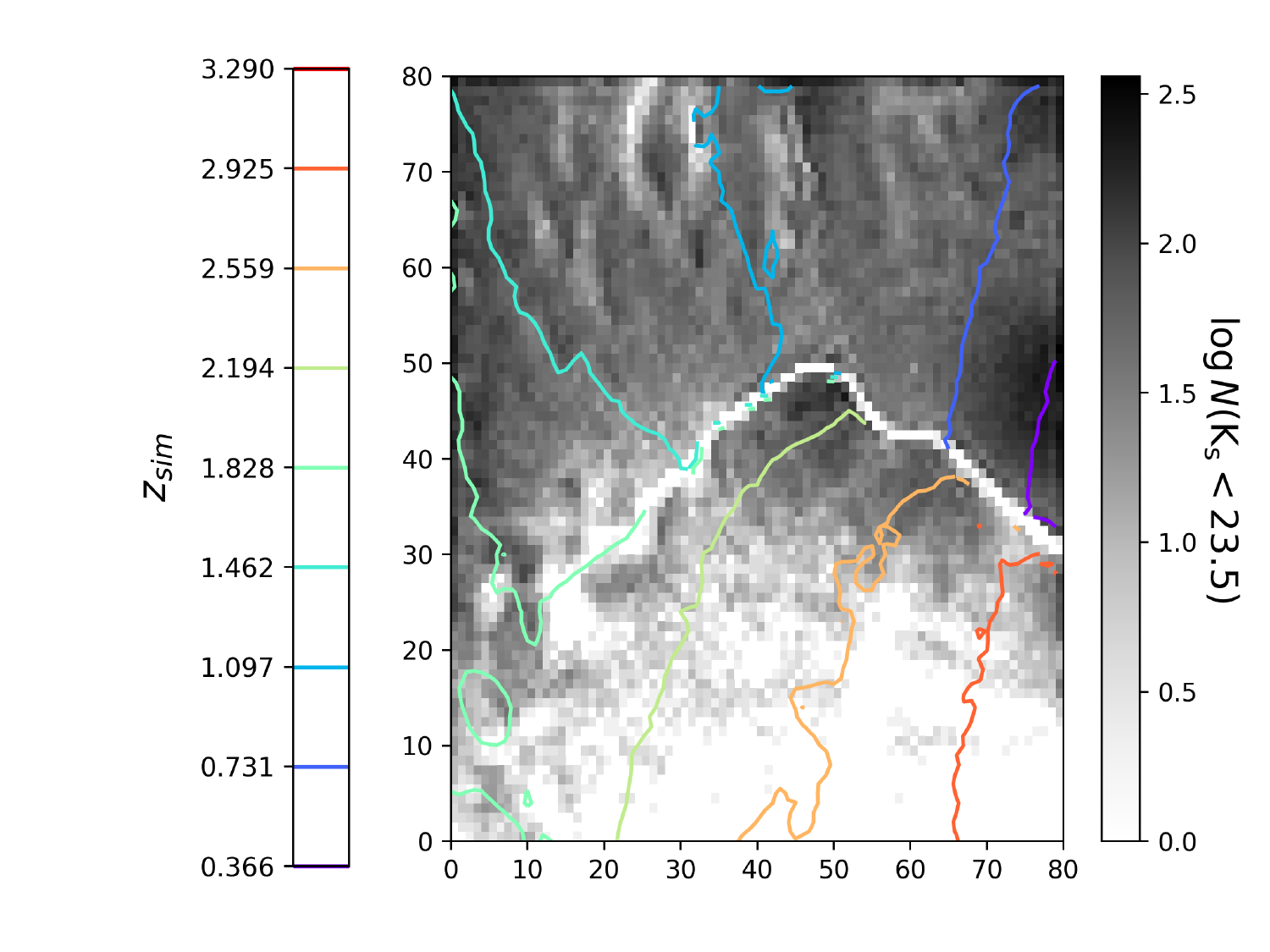} 

    \caption{The dark region represents the log of the number of galaxies projected on the SOM when $K_s < 23.5$ is applied to the dataset. The $K_s$-limited sample contains 232,118 objects.
    The rainbow contours represent levels of values of redshift (Figure\,\ref{figure4}), with redder colors representing higher redshift.}
    \label{figure5}
\end{figure}
\section{Results}
\label{sec:results}

Now that we have a trained and calibrated SOM, we use it to both explore the selection functions for our observations-like simulated dataset (Sec.\,\ref{subsec:visualization_obs}), and the derivation of redshift and stellar population parameters (Sec.\,\ref{subsection:parameter_estimation}). Specifically, for the latter we account for photometric errors (Sec.\,\ref{subsec:phot_err} and \ref{subsec:likelihood}) and cases of missing data (Sec.\,\ref{subsection:UL_MD}). 

\subsection{Visualization of Sample}
\label{subsec:visualization_obs}
One of the main strengths of SOMs is its ability to provide interpretable visualization of high-dimensional data. In particular, once a SOM is trained using the mass-limited dataset, it is possible to project on the map different samples derived by applying different cuts. 
Figure\,\ref{figure5} shows the projection of the $K_s$-limited sample, or HSC-Deep joint-like sample, reaffirming the conclusion of Figure\,\ref{figure1}. Indeed, a lack of objects at $z\gtrsim2.5$ with log($M_*/M_{\odot}$)$<10$ is present on the SOM. This is due to the stellar mass incompleteness introduced by the $K_s$ magnitude cut. Additionally, we notice that the ``missing" galaxies in the HSC-Deep joint-like sample are predominantly higher sSFR galaxies with log(sSFR)$\gtrsim$-9.2 yr$^{-1}$. In observed terms, these are galaxies with blue optical (e.g. $g-Y$ and $r-i$), but red near/mid-IR (e.g. $J-H$ and $ch1-ch2$) colors.
This projection of the $K_s$-limited sample highlights the effectiveness of SOMs in visualizing the impact of both magnitude and color cuts.

\begin{figure*}%[h]
\centering
    
     \includegraphics[width=1\textwidth,trim={3cm 5.5cm 0 6cm}]{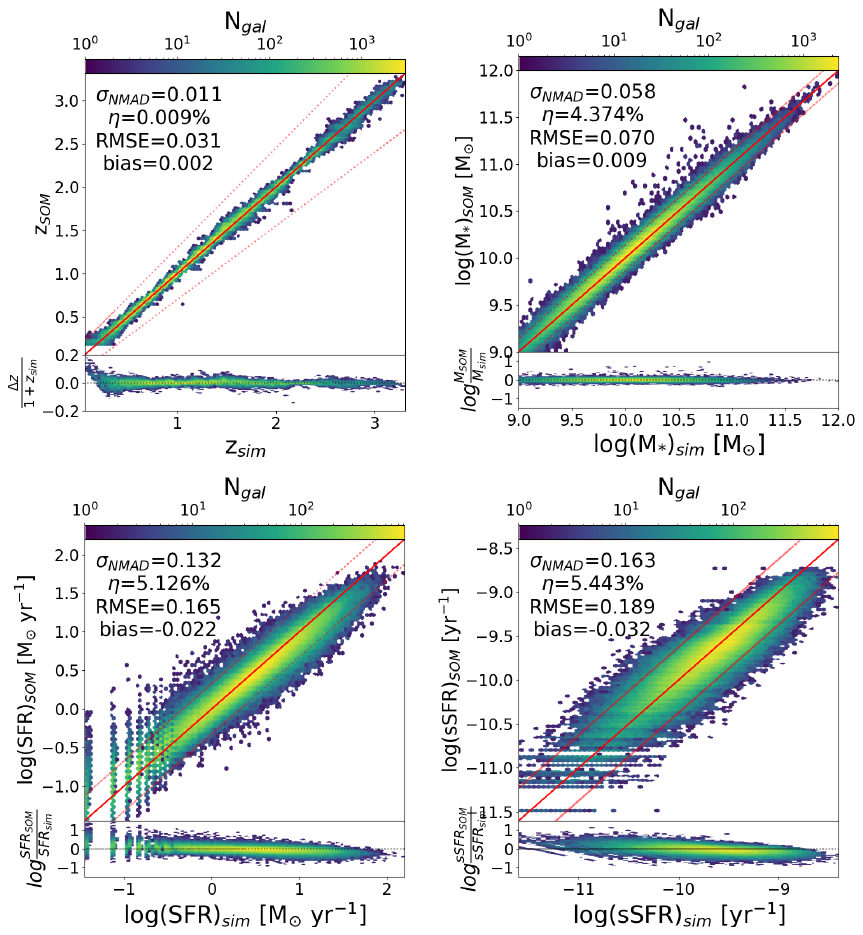} 

     \caption{The true simulated vs. SOM-derived parameter values in the ideal case with no photometric errors. These are based on the $K_s$-limited test dataset. The solid red line represents the bisection line; the red dashed lines indicate the boundaries of the outlier. The number of objects is 232,111.}
     \label{figure6}
\end{figure*}

\subsection{Deriving redshift and stellar population parameters from SOM} \label{subsection:parameter_estimation}

We test the use of SOMs in deriving redshift and stellar population parameters by projecting the $K_s$-limited dataset onto the SOM. For each galaxy, we assign median of $z$, log(nM$_*$), log(nSFR), and log(sSFR) of the SOM pixel in which it resides. To de-normalize M$_*$ and SFR, we use equation \ref{eq:normalization} and the observed $J$-band magnitude for each galaxy. Note that while we employ the $K_s$-limited sample for predictions, the SOM was trained and its pixels labeled with a different dataset, specifically the training datset, which was randomly drawn from the mass-limited dataset. Although the distributions of these two datasets in terms of $z$, M$_*$ and SFR may differ, we expect the mapping color space-physical parameter space to be consistent.

Figure\,\ref{figure6} shows the comparison between the true (in values from the HorizonAGN simulation) and the SOM-derived values for $z$, M$_*$, SFR and sSFR. For each parameter we also show four statistics including normalized median absolute deviation ($\sigma_{NMAD}$), outlier fraction $\eta$, root mean squared error (RMSE) and bias. The outlier fraction $\eta$ is the percentage of objects that satisfy $|z_{SOM} - z_{sim}|/(1+z_{sim}) > 0.15$ in the case of $z_{\rm phot}$, and $|logX_{SOM} - logX_{sim}| > 2\sigma$ in the case of the other properties $X$. Here $\sigma$ is the standard deviation of the difference log(predicted) - log(true).  

We find that redshift outperforms other statistics, with $\sigma_{NMAD}$ and RMSE of 0.01 and 0.03, respectively. Stellar mass estimation is also accurate, with $\sigma_{NMAD}$ and RMSE of 0.06 and 0.07, respectively. As typical of all parameter estimation methods, the estimation of SFR and sSFR has higher uncertainties. 
The spread between the true and SOM-derived parameters in Figure\,\ref{figure6} is due to the intrinsic spread $\delta$ in these parameters within a SOM pixel as shown in Figure\,\ref{figure4}. This spread does not include the effect of photometric errors. Therefore, the performance statistics presented here represents the ideal scenario.

\begin{figure*}%[H]
\centering
    \includegraphics[width=1\textwidth,trim={3cm 5.5cm 0 6cm}]{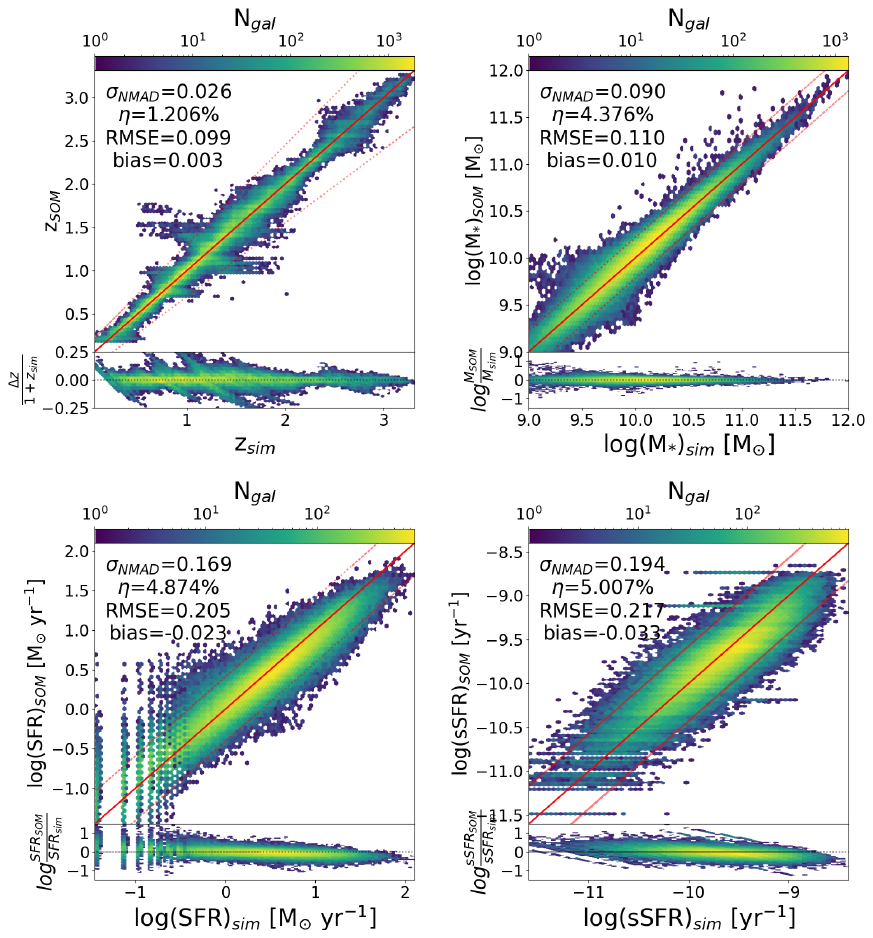} 

     \caption{The true simulated vs. SOM-derived parameter values in the more realistic case including photometric errors. These are based on the $K_s$-selected test dataset. The solid red line indicates the bisection line; the red dashed lines show the boundaries of the outlier. The number of objects is 227,365. Circa 2$\%$ of objects is lost because of invalid values from the log.}
     \label{figure7}
\end{figure*}

\begin{figure*}
    \centering
        \includegraphics[width=1\textwidth]{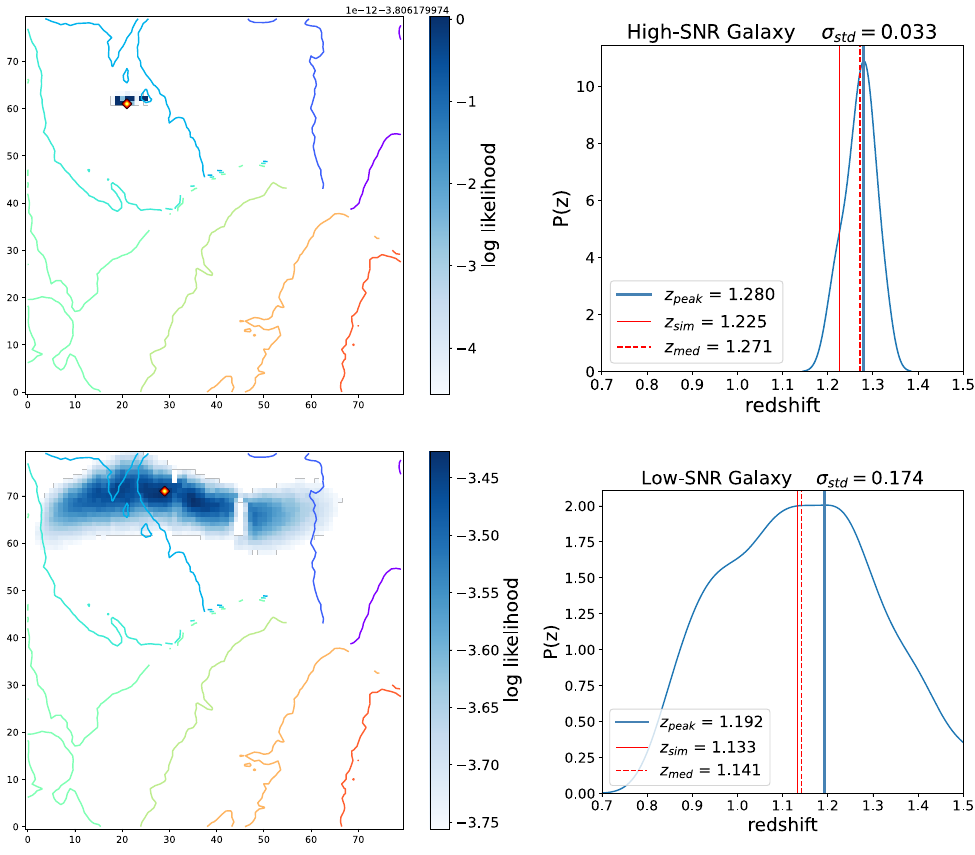} 

    \caption{The panels on the left show the normalized likelihood surfaces for two test galaxies: a high-SNR ($=82.3$ in $g$ band, $m_g$=23.4) example on top, and a low-SNR ($=4.0$ in $g$ band, $m_g$=27.7) example on the bottom both with true redshifts around $z\sim1.1-1.2$. The rainbow colors correspond to the redshift levels as in Figure.\,\ref{figure5}. The panels on the right show the redshift probability distribution based on these likelihood surfaces. The blue solid line represents the most likely redshift value, $z_{peak}$, from the 2D surface. The red solid line represents the true redshift, $z_{sim}$, while the red dotted line corresponds to the median of redshift, $z_{med}$, used to label the pixel in which the test galaxy is located. Note that the standard deviation $\sigma_{std}$ of the distribution is smaller for an higher-SNR test galaxy.}
    \label{fig:likelihood_surfaces}
\end{figure*}

\subsection{Accounting for photometric uncertainties}
\label{subsec:phot_err}

To consider photometric uncertainties, we adopt relative errors per magnitude derived from the HSC DR2 preliminary u2K catalog\footnote{\url{https://hscdata.mtk.nao.ac.jp/schema_browser3/##dr2.s18a_dud_u2k.forced}} for $u$ through $K_s$ bands \citep{Aihara_19}. We focused on the XMM-LSS field, since its IRAC $ch1$ and $ch2$ fluxes and uncertainties  were available from \citet{Krefting2020,Nyland2023}. 
We computed the relative photometric errors $\sigma_{\nu}$ in each band by fitting a line between log($\sigma_{\nu}/f_{\nu}$) and magnitude in that band. This approach accounts for larger relative errors for the more abundant faint objects and vice versa.  

For each galaxy in our simulated HSC-like dataset, we draw a new random flux value from a gaussian distribution centered at the true flux and a width given by the photometric error. 
We project these new perturbed data onto the SOM trained with noiseless colors. Afterwards, we repeat the procedure to estimate parameters as described in Section\,\ref{subsection:parameter_estimation}. Note that to de-normalised M$_*$ and SFR, we use the $J$-band magnitudes with added noise.  

Figure\,\ref{figure7} shows the comparison of true and SOM-derived parameters, accounting for photometric uncertainties. As expected, parameter uncertainties increase with photometric errors, but the estimations remain reasonably accurate. For instance, the new redshift $\sigma_{NMAD}$ is 0.026 with only a 1.2\% outlier fraction. The true vs.\,SOM-derived $z_{\rm phot}$ distribution in Figure\,\ref{figure7} however shows horizontal stripes where the $z_{\rm phot}$ derivation is less reliable, with increased number of outliers, at specific redshift ranges. The most evident stripe is at $z_{SOM}\sim 1-2$. We investigated the primary cause of these effects by examining the position on SOM of these outliers. The latter are all located near the area of the abrupt change in $K-ch2$ and $ch1-ch2$ (Figure\,\ref{figure3}), as a result of the move across the 1.6$\mu$m bump. The larger photometric errors in the IRAC channels can cause galaxies to fall into neighboring SOM pixels associated with different redshift values. We validated this hypothesis by repeating the same analysis of this section but removing the IRAC photometric errors. As a result, the horizontal stripes visible in the upper-right panel in Figure\,\ref{figure7} disappeared, confirming that the colors associated with the IRAC-channels were indeed the primary contributors.
This illustrates how photometric noise in specific bands impacts derived parameter uncertainties.

Due to the degeneracies between photometric redshift and shape of the SED, the outliers of the horizontal stripes are responsible also for the outliers in mass and SFR derivation visible in Figure\,\ref{figure7}. For example, if objects at somewhat higher-$z$ are incorrectly placed at lower-$z$ (right-side of the stripe), their redder colors are associated to more quiescent galaxies, leading to an underestimation of SFR, and vice-versa. Similarly, if an object is placed at a lower redshift than its true value, the observed $J$-band suggests a lower stellar mass than in reality. This results in stripes below the 1:1 relation in $M_*$ (see Figure\,\ref{figure7}). Since sSFR is the ratio of SFR and $M_*$, these effects are canceled out. This explains why our estimates for sSFR changed the least between the no-error and error case (Figure\,\ref{figure6} vs.\,Figure\,\ref{figure7}). We tested this interpretation by running the analysis without any IRAC channels errors, and as a result, the horizontal stripes in $z_{\rm phot}$, in $M_*$ ($logM_{*,SOM}/M_{\odot}\sim 9.5-10$), and in SFR (log(SFR)$_{SOM}\sim 0$) disappeared.

\subsection{Uncertainty estimation for individual galaxies}
\label{subsec:likelihood}

Figure\,\ref{figure7} gives an overall sense of parameter uncertainties when photometric errors are considered. These parameter uncertainties can be statistical, resulting from photometric errors, and systematic, resulting from parameter value distribution within a single SOM pixel. Below, we explain how we compute both statistical and systematic parameter uncertainties for individual galaxies.

For any galaxy, the likelihood $\mathcal{L}$  of it occupying a SOM pixel with coordinates [x,y] is computed as:
\begin{equation}
    \mathcal{L}_{[x,y]} = e^{-\frac{\chi^2_{[x,y]}}{2}}
    \label{eq:L}
\end{equation}
with:
\begin{equation}
    \chi^2_{[x,y]} = \sum_{i=1}^{10}\left(\frac{C_i - \overline{C_i}}{\sigma_{C_i}} \right)_{[x,y]}^2
    \label{eq:chi}
\end{equation}
where, in the pixel [x,y], $C_i$ are the 10 colors of the test galaxy, $\overline{C_i}$ are the 10 median colors in that pixel, and $\sigma_{C_i}$ represent the uncertainties of $C_i$. The likelihood surface is normalized by dividing $\mathcal{L}_{[x,y]}$ by its sum. Figure\,\ref{fig:likelihood_surfaces} (\textit{left panels}) shows the likelihood surfaces for two test galaxies. These test galaxies are of similar redshift ($z\sim1.2$) and were chosen to represent a high-SNR (signal to noise ratio) galaxy (pixel = [61, 21], SNR=82.3 in $g$ band, $m_g$=23.4) and a low SNR galaxy (pixel = [71, 29], SNR=4.0 in $g$ band, $m_g$=27.7). 
The likelihood surface of the high-SNR test galaxy is very concentrated around its true position on the SOM, while for the low-SNR test galaxy it is more spread out. 

From these likelihood surfaces, combined with the pixel labels for a given parameter, we can derive probability distribution functions for any of our parameters. For example, the right-hand panels of Figure\,\ref{fig:likelihood_surfaces} show the redshift probability distribution functions $P(z)$ derived from these likelihood surfaces.
The $P(z)$ of the low-SNR test galaxy exhibits a larger spread compared to the high-SNR test galaxy. This confirms the idea that the accuracy of parameter estimation with SOMs depends on a galaxy's SNRs, as indeed is the case for all methods used in deriving galaxy parameters. 

We define the statistical uncertainty for the given parameter, $\sigma_{stat,param}$, for any individual galaxy as the standard deviation of the probability distribution function, such as the $P(z)$ functions discussed above. To estimate the parameter systematic uncertainty $\sigma_{sys,param}$, we take half of the $\langle\delta\rangle=84\%-16\%$, the dispersion of galaxy parameter within the best-fit pixel for that galaxy (see right panels of Figure\,\ref{figure4}). The total uncertainty is:
\begin{equation}
	\sigma_{tot,param}=\sqrt{\sigma_{stat,param}^2+\sigma_{sys,param}^2}
    \label{eq:TOTerr}
\end{equation}
For example, to derive the uncertainties in the redshift estimation for the high-SNR test galaxy we look at its pixel [61,21] which has a redshift spread $\langle\delta\rangle=0.057$, hence $\sigma_{sys}=0.029$, leading to a total redshift uncertainty of $\sigma_{tot,z}=\sqrt{0.033^2+0.029^2}=0.044$. To compare this with the nominal $\sigma_{NMAD}$ in Figure\,\ref{figure7}, we need $\sigma_{tot,z}/(1+1.2)=0.02$, which is consistent with the $\sigma_{NMAD}$ of 0.026.

\subsection{Handling missing data}
\label{subsection:UL_MD}

\begin{figure*}
\centering
     \includegraphics[width=0.9\textwidth]{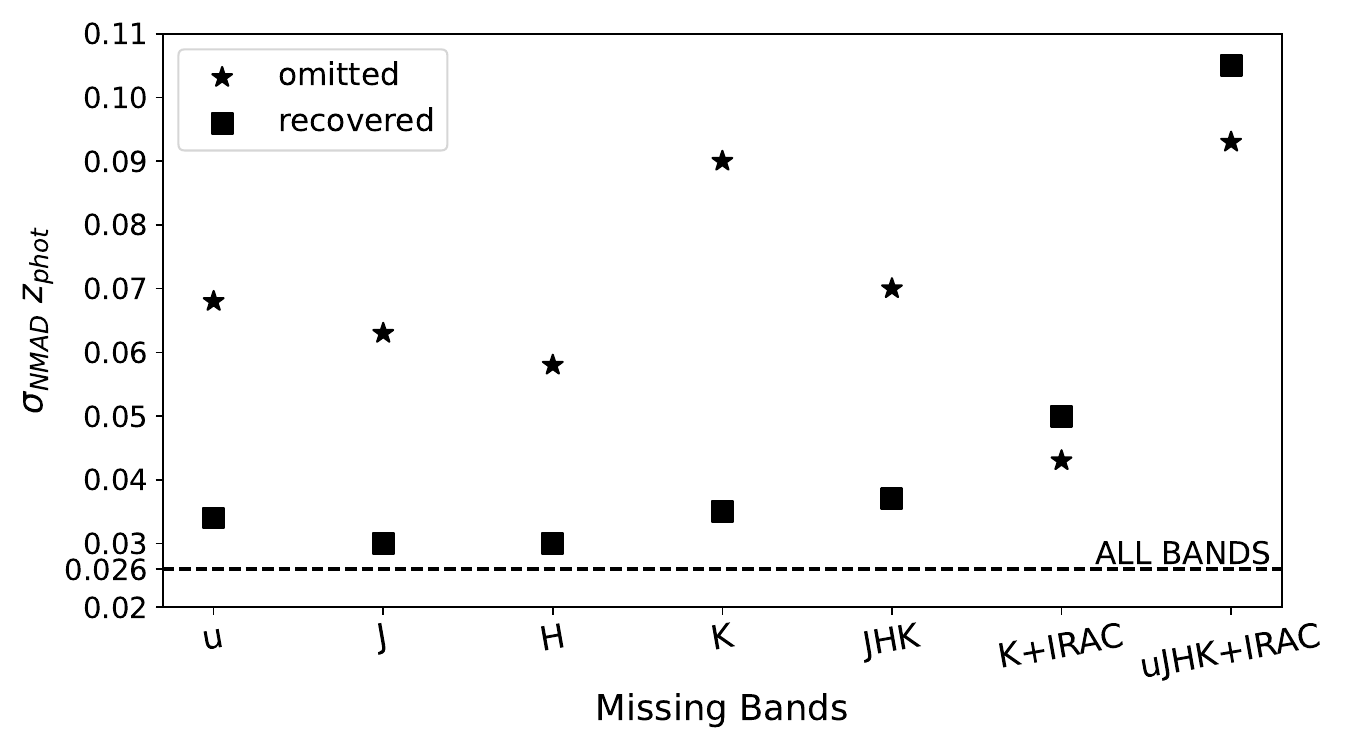} 
     \caption{The $\sigma_{NMAD}$ for estimated $z_{\rm phot}$ as a function of missing bands. Stars represent the case when the missing band(s) are omitted from the SOM training. Squares represent the case when the SOM is trained on the complete data and datasets with missing data are projected onto it using the missing color recovery method described in Section\,\ref{subsection:UL_MD}. The horizontal dashed line indicates the $\sigma_{NMAD}=0.026$ achieved using all bands (see Figure\,\ref{figure7}).}
    \label{figure:snmad}
\end{figure*}

The previous section's analysis assumes flux measurements are available across all bands. In reality, the surveys making up the HSC-Deep joint dataset do not perfectly overlap. Several square degrees of the HSC-Deep joint survey in fact will have no data outside the core bands of $grizY$. Additionally, even in overlapping areas, there can be instances of missing data in individual bands, due to issues such as bad data arising from image artifacts. The simplest approach would be to train a SOM excluding any missing bands. We tested this scenario for multiple individual missing bands or sets of missing bands. 
Figure\,\ref{figure:snmad} shows the $z_{\rm phot}$ $\sigma_{NMAD}$ in case of missing $u$-band, $J$-band, $H$-band, $K_s$-band, $JHK_s$, $K_s$+IRAC, and $uJHK$+IRAC. The $\sigma_{NMAD}$ gets worse compared to the baseline of complete data (dashed line) with $\sigma_{NMAD}$=0.026. Note that this analysis takes into account photometric uncertainties in the same manner as done in Section\,\ref{subsec:phot_err}, where each flux is perturbed by Gaussian noise based on the given flux uncertainty.

The degradation in parameter estimation seen in Figure\,\ref{figure:snmad} is due to the loss of information resulting from the missing bands. To address this, we assume that galaxies with missing data fall within the same color distribution as the training sample. This is justifiable when the missing data is simply due to the inhomogeneous coverage of the various multiband surveys. To recover the colors involved with missing bands, we perform 500 random draws from the color distribution of these same colors, as defined in the training set. Our color distribution again include photometric uncertainties as discussed above. The random draws involve either 1D, 2D or 3D color distributions depending on whether the missing band is on the edge (e.g. $u$-band), in the middle (e.g. $H$-band), or a set of adjacent bands is missing (e.g. $JHK$). With our 500 random draws for the ``missing band" colors combined with the existing colors for each galaxy, we are able to essentially build a SOM likelihood surface for that galaxy in a manner similar to that discussed in Section\,\ref{subsec:likelihood}. The adopted parameter value is then simply based on the maximum-likelihood SOM pixel. The resulting redshift $\sigma_{NMAD}$ values for each of our missing data scenarios are shown in Figure\,\ref{figure:snmad}. We note that in most cases, this band recovery procedure allows us to achieve significantly better $\sigma_{NMAD}$ values than by simply training the SOM without the missing data. We emphasize that this improvement is based on the assumption that the color distributions determined from the data with complete coverage are representative of the data with incomplete coverage as well. 

The only cases where our recovered data perform worse than the SOM trained by omitting the missing data are those involving IRAC channels. We tested that this worsening does not depend on the adopted photometric error. This effect is likely due to a more uncertain extrapolation when more than one color needs to be recovered right on the edge of the wavelength coverage as is the case of missing the IRAC channels.

\begin{table}[h]
\centering
\caption{Effect of parameter recovery for different missing bands ``filled-in" as described in Section\,\ref{subsection:UL_MD}.}
\begin{tabular}{l|ccc}
 & \multicolumn{3}{c}{\textbf{$\sigma_{NMAD}$}} \\ \hline %& \multicolumn{3}{c}{\textbf{RMSE}} \\ \hline
\textbf{All Bands} & $z_{\rm phot}$ & $M_*$ & $SFR$ \\ \hline %& z & M & SFR \\ \hline
 & 0.026 & 0.090 & 0.169 \\
 \hline
 \hline
\textbf{Missing Bands} &$z_{\rm phot}$ & $M_*$ & $SFR$ \\ \hline %& z & M & SFR \\ \hline
u & 0.034 & 0.102 & 0.240 \\
uJ/uJH & 0.041 & 0.158 & 0.276 \\
uJK/uJHK & 0.056 & 0.224 & 0.321 \\
uJHK+IRAC/uJK+IRAC & 0.105 & 0.336 & 0.484 \\
uH & 0.040 & 0.145 & 0.279 \\
uHK & 0.053 & 0.244 & 0.336 \\
uHK+IRAC & 0.090 & 0.318 & 0.480 \\
uK & 0.043 & 0.156 & 0.278 \\
uKIRAC & 0.068 & 0.290 & 0.383\\
J/JH & 0.030 & 0.141 & 0.208 \\
JHK & 0.037 & 0.172 & 0.226 \\
JK+IRAC/JHK+IRAC & 0.071 & 0.305 & 0.399\\
JK & 0.037 & 0.171 & 0.226\\
H & 0.030 & 0.127 & 0.202\\
HK & 0.042 & 0.181 & 0.218 \\
HK+IRAC & 0.066 & 0.292 & 0.349\\
K & 0.035 & 0.144 & 0.189\\
IRAC-\textit{ch1} & 0.035 & 0.117 & 0.208\\
K+IRAC & 0.050 & 0.237 & 0.304\\ \hline
\end{tabular}
\label{tab:snmad_parameters}

\end{table}

\subsection{Parameter recovery vs. wavelength coverage}

The missing data analysis also provides us with information on the relative importance of having particular bands present for the recovery of our parameters. Table\,\ref{tab:snmad_parameters} goes beyond $z_{phot}$, shown in Figure\,\ref{figure:snmad}, to explore how the $\sigma_{NMAD}$ for $z_{phot}$, $M_*$ and $SFR$ all vary depending on which bands are missing and filled in with our recovery method (Section\,\ref{subsection:UL_MD}). 
We reach reasonably good accuracy for when only a single band is missing. 
As more bands are missing and filled-in, the values of $\sigma_{NMAD}$ increase. For example, the worst parameter estimation occurs when all ancillary bands outside the HSC-Deep core bands (\textit{grizY}), namely $uJHK$+IRAC, are missing and filled-in with our method. If at least $u$-band, among the ancillary bands, is not missing (i.\,e.\,in the case of recovered missing $JHK$+IRAC) the parameter $\sigma_{NMAD}$ improves especially for $z_{\rm phot}$ estimation, with a decreasing in $\sigma_{NMAD}$ of $\sim 32\%$.
These findings show the critical importance of the ancillary surveys providing the $u$-band, the near-IR, and the \textit{Spitzer} IRAC $ch1$ and $ch2$ coverage.

\begin{figure*}
\centering
     \includegraphics[width=1.2\textwidth,trim={7cm 1cm 0 0cm}]{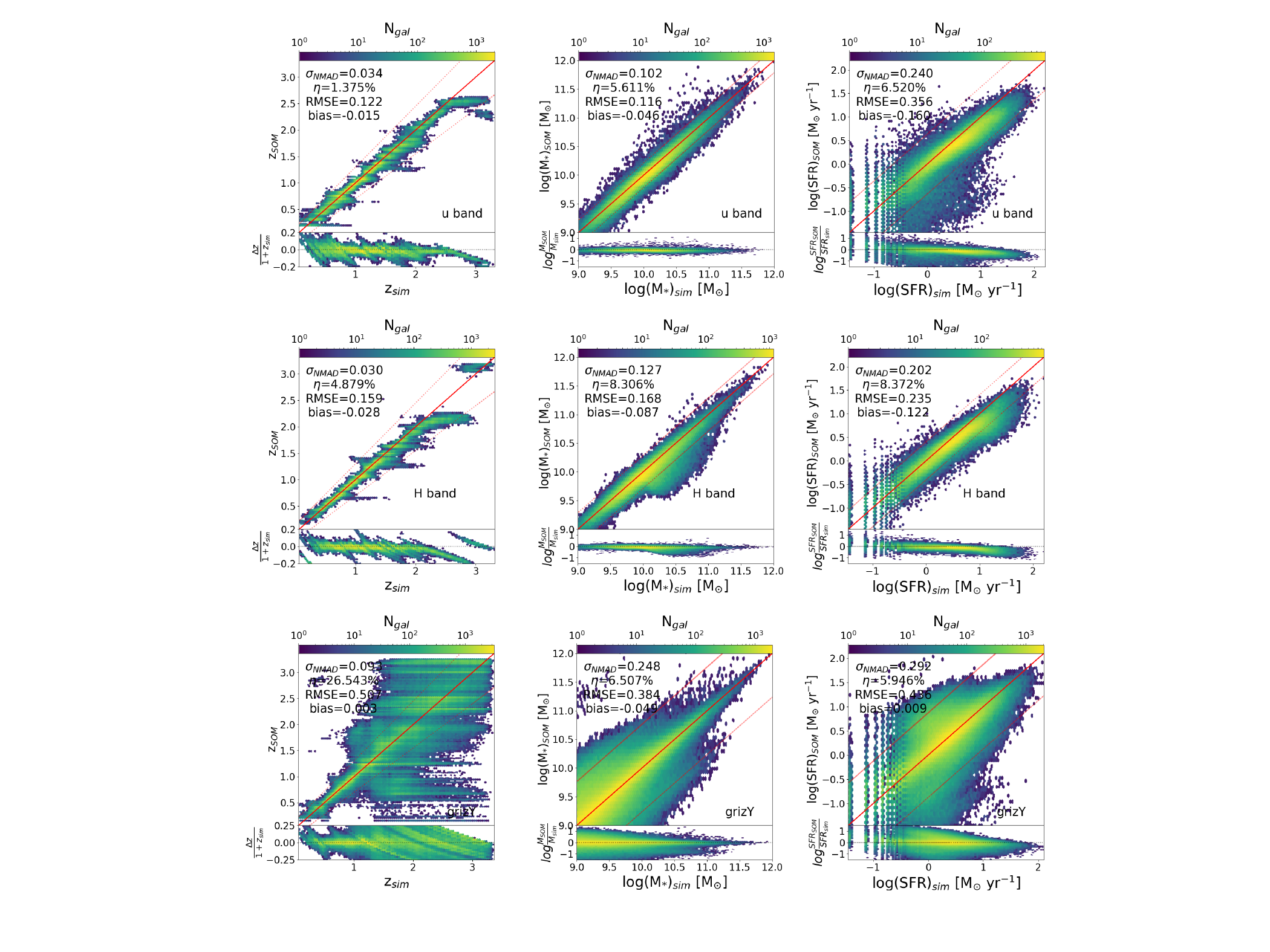} 
     \caption{From left to right, $z_{\rm phot}$, $M_*$, and SFR predicted vs. true in case of recovered missing data for $u$-band (\textit{top}) and $H$-band (\textit{middle}). The procedure for handling missing data is described in Section\,\ref{subsection:UL_MD}. \textit{Bottom}: predicted vs. true in case only HSC-Deep $grizY$ bands are available and the missing bands are not recovered. Red lines as in Figures\,\ref{figure6},\ref{figure7}.}
    \label{fig:missbands_param}
\end{figure*}

The variation of parameter $\sigma_{NMAD}$ with specific bands depends on their different sensitivity to galaxy parameters at different ranges.
Figure\,\ref{fig:missbands_param} illustrates this for the cases when $u$-band (top panel), $H$-band (middle panel), and all ancillary to HSC-Deep data are missing (bottom panel). 
We chose these specific cases as the $u$-band is crucial for accurate $z_{\rm phot}$ estimation at $z\gtrsim2.5$ (as this filter enters the Lyman break) and traces the rest-frame ultraviolet hence is relevant for the SFR estimation. The top panel in Figure\,\ref{fig:missbands_param} shows that indeed when the $u$-band is missing our redshift recovery gets worse beyond $z\sim2.5$, the SFR recovery is poor. As seen in Table\,\ref{tab:snmad_parameters}, the SFR recovery here is the worst compared to any other single (or couple of) missing band cases. 

The $H$-band, in our redshift range, traces the rest-frame optical to near-IR, therefore it is sensitive to the stellar mass. It also affects the $z_{\rm phot}$ determination since that is sensitive to the 1.6\,$\mu$m bump at lower $z$, and to the 4000$\AA$-break at $z > 3$. The middle panel in Figure\,\ref{fig:missbands_param} shows that indeed, when the $H$-band is missing our redshift recovery is much worse at $z>2$. The underestimated redshifts in this regime also lead to a pronounced bulge of underestimated stellar masses, and corresponding underestimated SFRs. 

The last case we explored in more detail was the worse case scenario (see Table\,\ref{tab:snmad_parameters}) which was the missing $uJHK+IRAC$ scenario, or when we have no data outside the HSC-Deep core bands ($grizY$). In this case, we use the scenario where the missing data are simply omitted rather than recovered, since the later shows worse parameter recovery (see Figure\,\ref{figure:snmad}). We see in the bottom panel of Figure\,\ref{fig:missbands_param}, that the redshift estimation is still reasonable at $z<1-1.5$, where the optical bands trace the 4000\,$\AA$ break, but there is essentially no handle on the redshift beyond that (up to the $z\sim3$ regime explored here). The longer wavelength ($J$+) bands are critical for the $z\sim1.5-3$ estimation due to their sampling both of the Balmer break and the 1.6\,$\mu$m stellar bump.

\subsection{Handling upper limits}
\label{sec:upper limits}
Besides cases of missing data, we can also have upper limits if the source is too faint relative to the noise in a given image. The naive first approach to handle such upper limits is to draw random fluxes from uniform distributions representing the upper limit. However, this method can lead to unrealistic colors that place the galaxy outside the SOM surface, which only represents the 10D color space present in the training set. Dealing with upper limits is in general a significant challenge for machine learning methods, as it requires extrapolating outside the training set. 

\begin{figure}
\centering
     \includegraphics[width=0.55\textwidth,trim={4.7cm 1cm 0 0.5cm}]{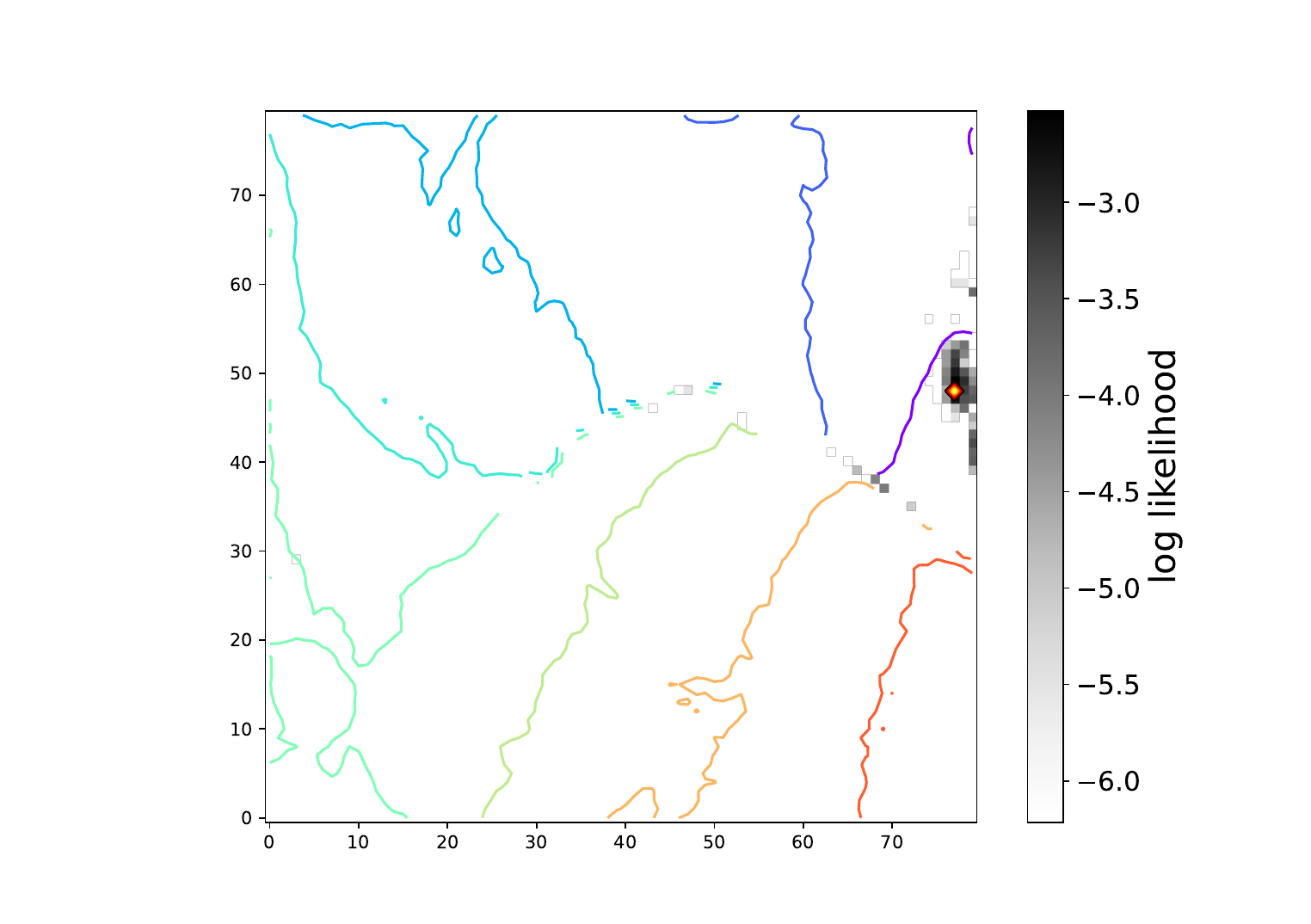} 
     \caption{The log of the normalized likelihood for a test galaxy (hot diamond) with an upper limit for $H$-band, in the scenario where the SOM is trained with complete deeper data. The rainbow contour represent levels of redshift values as in Figure\,\ref{figure5}.}
    \label{figure:upper_limit}
\end{figure}

While a more comprehensive treatment of upper limits in SOMs is beyond the scope of the present work, we can adapt our missing data approach for one particular scenario involving upper limits. This is the case where upper limits are a result of some part of the survey having significantly shallower coverage in some band relative to another deeper portion of the survey. In this case, we can train the model using the full color set from the deep portion of the survey. The assumption is that the deeper-data training set contains realistic colors for sources with upper limits in the shallower portion. In this case, we can implement a modification of the missing data approach described in the previous section. Instead of drawing from the full color distribution, we draw from a truncated color distribution that accounts for the detection limit in the shallower data. Figure\,\ref{figure:upper_limit} illustrates this approach with an example of a test galaxy having only an upper limit in $H$ band. The test galaxy is drawn from a fiducial dataset with $H-$band coverage two orders of magnitude shallower coverage than in the training set. In this case, the recovered colors are realistic and fall within the panchromatic space used to train our SOM, as showed by the grey region in Figure\,\ref{figure:upper_limit}. Note that the tightness of the SOM likelihood survey of this upper limit galaxy is due to the fact that each SOM pixel involves 10 colors which helps compensate to a large degree for any missing/upper limit data.

\section{Discussion}
\label{sec:discussion}
\subsection{Comparison with Davidzon et al}

This paper was originally inspired by \citet{D19}, who also explored the derivation of $z_{\rm phot}$ and SFR using \textsc{SomPY} on the Horizon-AGN simulated lightcone. They added photometric errors corresponding to the COSMOS photometry. Their training color-set excluded $g$-band colors and colors from non-adjacent bands (e.g. $g-Y$, $r-z$). In this paper, we expand on this work, first by applying it to a different photometric set (the HSC-Deep ones); exploring the effect of different color choices for training the SOM, and exploring the effects of missing data and how to handle them. Fundamentally, we achieve a photo-$z$ $\sigma_{NMAD}$ of 0.026 and an outlier fraction of 1.2$\%$, while \cite{D19} reported a $\sigma_{NMAD}$ of 0.044 and an outlier fraction of 6.1$\%$. While the approach is very similar, this is a significant improvement in performance. Our tests suggests that the key aspect behind this improvement is the specific choice of colors. We use more widely separated colors, such as $r-z$, which is sensitive to the redshift, and we also add the $g$-band (which is omitted in \citet{D19}) which also improves performance, especially in redshift regimes where it helps pinpoint the Balmer break. In addition, our improved performance is aided by our uncertainties in the $u$-band are $\approx0.2$mag lower, they are comparable in the visible bands, and $\approx0.1$mag lower in the NIR, although our IRAC channels errors are worse by $\approx0.2$mag. \citet{D19} do not quote detailed statistics for SFR, but they quote $\sigma_{NMAD}<0.2$ which is consistent with our results.

In a more recent paper, \cite{Davidzon_2022} used their SOM method to derive $M_*$ from the observed COSMOS data and compare their SOM-derived values with traditional SED-fitting. Their results indicates that this method works equally well as traditional SED fitting but it is computationally much faster, making it suitable for large datasets. Their distribution of SOM-derived vs.\,SED fitting-derived stellar masses appears visually wider than our SOM-derived vs.\,true stellar mass. However, this is not a fair comparison since there are uncertainties associated with the SED-fitting derived values as well.

\subsection{The handling of missing data}
In this paper, we explore for the first time in the astronomical context the handling of missing data in SOMs. This issue has been explored in SED template fitting codes, e.g.\,by modifying the $\chi^2$ method, usually at the expense of increased computational cost \citep{Sawicki_2012, cigale}. A recent paper, \cite{rejeb2023selforganizing} does discuss how to train SOMs with incomplete data; however, this paper did not appear until the final stages of our current work therefore we have not had a chance to implement it. In future works, we plan to explore the possible use of their algorithm through their publicly available code\footnote{\url{https://CRAN.R-project.org/package=missSOM}}. One needs to be cautious, however, about including incomplete data in the training, especially in cases of large variance, as discussed in \cite{Cottrell}. They argue that when a sufficient amount of data is available to build a robust model, it is preferable to train on the complete data and only afterwards consider the effects of missing data. Indeed, this is the approach we adopted in our study. In our paper, we outline a method to train the SOM on the complete dataset and ``recover" the missing data where the data are incomplete by assuming the SOM training set is otherwise representative of the galaxies with missing bands. 
In line with our finding that it is better to fill-in the missing data rather than simply omit it, \citet{Chartab_2023} achieved better results in SED fitting after first using a random forest model to fill-in any missing data.

\section{Summary \& Conclusions}

\label{sec:conclusions}
SOMs are an efficient means of grouping galaxies based on their colors, providing a powerful means for visualizing multi-dimensional galaxy datasets as well as estimate any SED-based parameters. In this study, 
we train a SOM on a mass-limited light-cone from the Horizon-AGN simulation that has been color-calibrated on real data. Our training color set is based on a fiducial $ugrizYJHK_s$+IRAC photometric coverage characteristic of the HSC-Deep survey plus ancillary data (the HSC-Deep joint survey).  We also examine an observations-like magnitude-limited sample with $K_s<23.5$. We assess the quality of the SOM-derived photometric redshifts, stellar masses, SFRs and sSFRs while considering realistic photometric errors for the HSC-Deep joint survey. Finally, we investigate the effects on the derived parameters in cases of missing data. 
Our key findings are:

\begin{itemize}
    \item 
    The set of colors used for SOM training should be those most sensitive to the parameters we intend to estimate. For example $u-g$, $J-H$ and $ch1-ch2$ are all colors that are critical to accurate redshift estimation, each particularly sensitive to different redshift ranges. 
    \item SOMs can be used to interpret visualization of high-dimensional data. For example, the projection of the $K_s$-limited sample onto the SOM (Figure\,\ref{figure5}) shows that this selection is similar to the mass-limited sample up to $z\sim1$, but loses galaxies with higher sSFR (log(sSFR)$>$-9) and lower stellar mass (log($M_*/M_{\odot}$)$<$10), beyond that. 
    \item With the HSC-Deep joint survey photometric uncertainties, the derived $z_{\rm phot}$'s have $\sigma_{NMAD}=0.026$. For M$_*$, SFR, and sSFR 
    we have $\sigma_{NMAD}<0.2$. 
    The IRAC channels uncertainties are particularly critical in the $z\sim1-2$ range where the $K-ch2$ and $ch1-ch2$ colors trace the 1.6$\mu$m bump. 
    \item We consider both the systematic and statistical uncertainties in the derived parameters for individual galaxies. 
    In particular, the probability distribution function is derived from the likelihood surface of a galaxy within the SOM, which is strongly dependent on the signal-to-noise ratio of the particular galaxy. 
    \item We discuss how to recover and project cases of missing photometric bands onto a SOM trained with the full dataset. This leads to improved parameter estimations compared to simply omitting missing data from the training. In limited cases, this approach can be applied to upper limits as well.
    \item The quality of the derived parameters is strongly dependent on the overall wavelength coverage included in training the SOM. We provide a table for comparison of achieved $\sigma_{NMAD}$ for estimated $z_{\rm phot}$, $M_*$, and SFR when different (sets of) missing bands are recovered.

\end{itemize}

We stress out that the findings presented in this study are based on simulations, despite our efforts to make the simulations more realistic through the color-calibration process. Our next step will be to use the methods outlined here to use SOMs for parameter estimation for the upcoming HSC-Deep joint catalog. Finally, in a companion paper (La Torre et al. 2024b in prep.), we conduct a performance comparison of SOM with three supervised ML methods: extreme gradient boosting, fully connected neural networks, and random forest. For comprehensive information on the training process and our findings, we refer the reader to that paper.

\appendix
\twocolumngrid
\section{Comparison of SOMs Trained with Simulated and Observed Datasets}
\label{appA}
\begin{figure}
\centering
    \includegraphics[trim={0.5cm 9.3cm 0cm 0cm},width=0.81\textwidth]{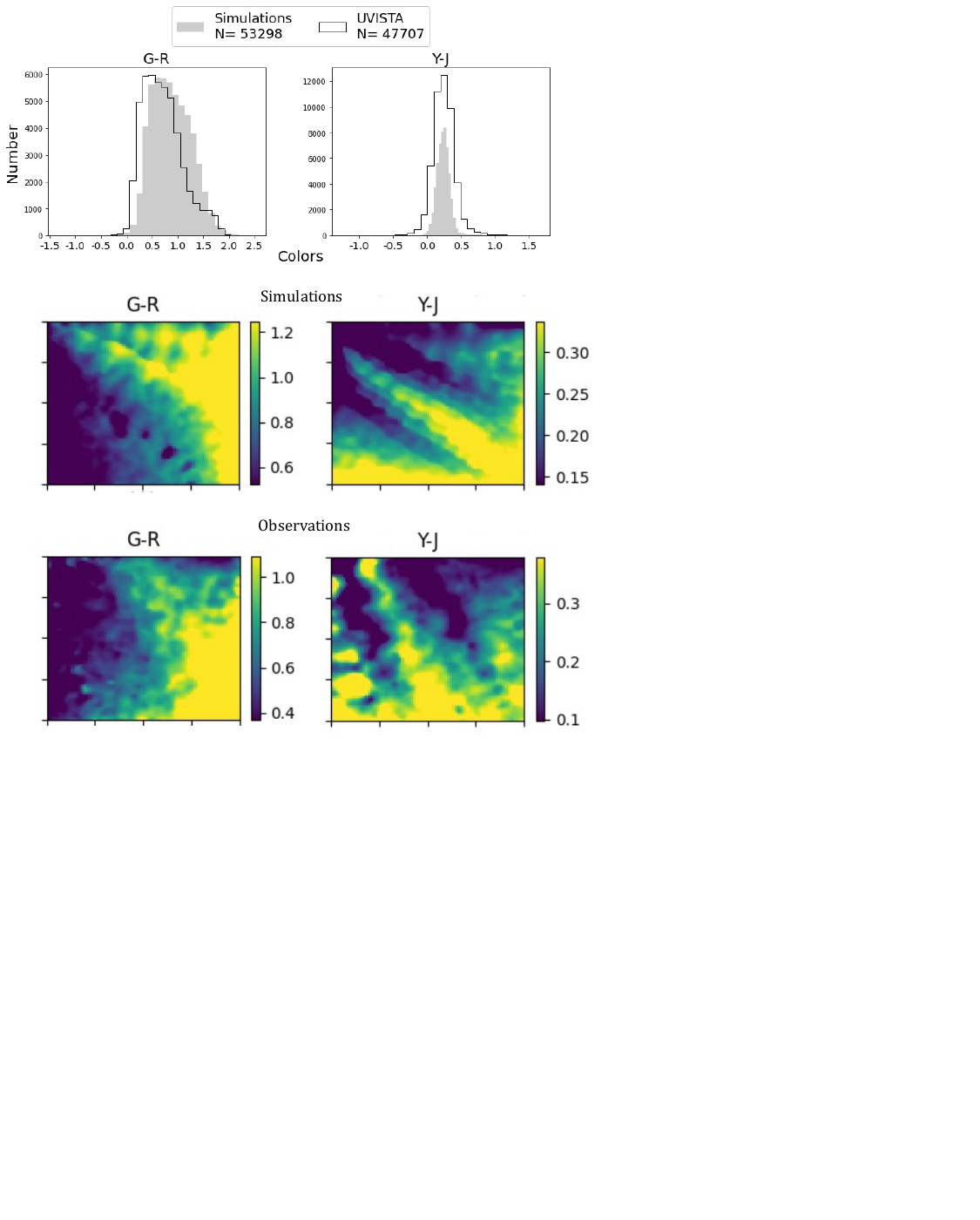}

    \caption{The top panel shows the comparison of distribution of simulated and UVISTA colors, obtained from magnitudes brighter than the peak of the observed magnitude's distribution in each band. The colors \textit{g-r} on the left and \textit{Y-J} on the right are shown as an example of two similar color distributions. 
    The middle and bottom panels show the comparison of component maps obtained from training SOM with simulated galaxy colors (middle) and UVISTA (bottom). A comparison between the middle and the bottom panels shows that the two datasets yield the same component maps, meaning that if the input objects have the same color distribution, as in the top panel, than even the SOMs will be similar. For this test we trained the two SOMs with the same color set considered in this paper.}
    \label{figure10}
\end{figure}

We test if two SOMs trained with a simulated or an observed dataset yield similar component maps. This is important as our final goal is to employ SOM to estimate galaxy parameters for the large HSC Deep joint survey. For this test we compare two SOMs, one trained with the mass-limited simulation and the other trained with observed galaxies from UltraVISTA catalog \citep{muzzin13}. The training colors of these two datasets should be similar if we want to check that simulated and observed galaxies with same colors are placed in same regions of the map after the training. To this effect, for this test we select only simulated and observed galaxies brighter than the peak of the observed magnitude's distribution in each band. Thus, the resulting simulated and observed color distributions are similar, as show in the top panel of Figure\,\ref{figure10}, and so they can be used to train two SOMs. Note that these components maps differs from Fig.\,\ref{figure3}, due to variations in the magnitude distribution utilized for color computation. In the analysis of this work, we exclusively apply $K_s$-magnitude cuts, whereas in this test, the cuts are made at the peak of the observed magnitude distributions. The middle and bottom panels in Figure\,\ref{figure10} illustrate the two SOMs color-coded by two of the training colors, showing that they yield similar component maps (more details in the figure caption). This means that if the two training datasets have same color distributions, than even the SOMs will be similar. Consequently, with the aim of using this analysis with similar observed color distributions, we can trust, within some limits, the analysis we are performing in this work with simulations.

\section*{Acknowledgments}

AS is grateful for fruitful discussions with Peter Capak, Iari Davidzon, and Shoubaneh Hemmati during a sabbatical leave at Caltech in 2019 which originally brought SOMs and their use in parameter estimation to her attention.  
We are grateful to Jenny Greene, ChangHoon Hanh, and Tom Loredo for helpful discussions. 
This work is support by NASA under award number 80NSSC21K0630, issued through the Astrophysics Data Analysis Program (ADAP). LSJ acknowledges the support from the Brazilian agencies CNPq (308994/2021-3) and FAPESP (2011/51680-6). The authors acknowledge the Tufts University High Performance Computing Cluster (\url{https://it.tufts.edu/high-performance-computing}) which was utilized for the research reported in this paper. The authors are grateful to the anonymous reviewers for their insightful feedback and constructive comments, which contributed to the improvement of this paper.

\software{Astropy \citep{astropy:2013, astropy:2018, astropy:2022}, \textsc{SomPY}\citep{moosavi2014sompy}.}
\bibliography{references}{}
\bibliographystyle{aasjournal}

\end{document}